\documentclass[%
 reprint,
 amsmath,amssymb,
 aps,
]{revtex4-2}

\usepackage{graphicx}
\usepackage{dcolumn}
\usepackage{bm}
\usepackage[hidelinks]{hyperref}
\usepackage{cleveref}
\usepackage{siunitx}

\usepackage{xcolor}

\usepackage{acronym}
\newacro{PSD}[PSD]{Power Spectral Density}
\newacro{SNR}[SNR]{Signal-to-Noise Ratio}
\newacro{GW}[GW]{Gravitational Wave}
\newacro{IMBH}[IMBH]{Intermediate-Mass Black Hole}
\newacro{SBI}[SBI]{Simulation-Based Inference}
\newacro{LVK}[LVK]{LIGO-Virgo-KAGRA}
\newacro{FMPE}[FMPE]{Flow Matching Posterior Estimation}
\newacro{SNPE}[SNPE]{Sequential Neural Posterior Estimation}
\newacro{NPE}[NPE]{Neural Posterior Estimation}
\newacro{NPSE}[NPSE]{Neural Posterior Score Estimation}
\newacro{SVD}[SVD]{Singular Value Decompositions}
\newacro{LAMPE}[LAMPE]{Likelihood-free AMortized Posterior Estimation}
\newacro{JSD}[JSD]{Jensen-Shannon Divergence}
\newacro{PDF}[PDF]{Probability Density Function}

\begin{document}

\preprint{APS/123-QED}

\title{Simulation-based Inference for Gravitational-waves from Intermediate-Mass Binary Black Holes in Real Noise}

\author{Vivien Raymond}
 \email{raymondv@cardiff.ac.uk}
\author{Sama Al-Shammari}
\author{Alexandre G{\"o}ttel}
\affiliation{%
 School of Physics and Astronomy, Cardiff University, Cardiff, CF24 3AA, United Kingdom
}%

\date{\today}

\begin{abstract}
We present an exploratory investigation into using Simulation-based Inference techniques, specifically Flow-Matching Posterior Estimation, to construct a posterior density estimator trained using real gravitational-wave detector noise. Our prototype estimator is trained on a 9-dimensional space, and for training efficiency outputs posterior probability distributions for the binary black holes chirp mass and mass ratio. We use this prototype estimator to investigate possible effects on parameter estimation for Intermediate-Mass Binary Black Holes, and show statistically significant reduction in measurement bias. Although the results show potential for improved measurements, they also highlight the need for further work.
\end{abstract}

\maketitle


\section{\label{sec:intro}Introduction}

The LIGO~\cite{PhysRevD.102.062003}, Virgo~\cite{PhysRevLett.123.231108} and KAGRA~\cite{10.1093/ptep/ptaa125} \ac{GW} observatories have since 2014 reported around one hundred \ac{GW}s from binary black hole mergers, with individual black hole masses ranging from a few stellar masses to over a hundred~\cite{gwtc3}. The heaviest detected so far lie at the edge of the regime associated with \ac{IMBH}, above about one hundred solar masses. As outlined in~\cite{annurev:/content/journals/10.1146/annurev-astro-032620-021835}, these black holes could play a fundamental role in stellar and galactic evolution, and represent a significant source of gravitational waves and tidal disruption events.
However, in the detection band of current ground-based gravitational-wave detectors, \ac{IMBH} signals are typically only detectable in very short time intervals of a few milliseconds. Thus, short duration noise artefacts (known as glitches) present in the noise of those detectors have the potential to strongly affect the measurements of \ac{IMBH}s. During LIGO's third observing run, glitches with a \ac{SNR} of at least 6.5 occurred about once per minute~\cite{gwtc3}, with the actual rate of relevant glitches being higher due to the potential impact of smaller glitches on parameter estimation.

The parameter estimation methods commonly used by the \ac{LVK} collaboration, such as those available in the Bilby library~\cite{ashton2019bilby}, have proven very reliable for inferring black-hole parameters. However, they generally assume that the noise around events is purely Gaussian and stationary. When glitches or other deviations occur, these assumptions can significantly impact the results~\cite{powell2018,davis2022,PhysRevD.106.042006,PhysRevD.105.103021,10.1093/mnras/stad341}. 
\ac{SBI} on the other hand, leverages machine learning to infer parameters without assuming a specific noise distribution, provided sufficient training data are available~\cite{cranmer2020frontier}. Previous studies have applied \ac{SBI} and \ac{NPE} towards \ac{GW} parameter estimation~\cite{green2020gravitational,delaunoy2020lightning,dax2021real,dax2022group,dax2023flow,Bhardwaj:2023xph,alvey2024simulation} using Gaussian stationary noise. In particular, ~\cite{dax2023flow} used \ac{FMPE} on gravitational-wave signals in Gaussian noise. The work presented here applied the same technique, on a more limited gravitational-wave signal space, but looking at the impact of real, non-Gaussian and non-stationary detector noise.
Notably,~\cite{wildberger2023adapting} has been able to address \ac{PSD} uncertainties, while~\cite{legin2023framework,2023mla..confE..17L} employ Score-Based Likelihood Characterization to create a likelihood function based on real detector noise.

In this work, we use \ac{SBI} to directly map simulated \ac{IMBH} signals in real detector noise to the posterior distributions of inferred black hole parameters, for training efficiency we currently only infer the binary's chirp mass and mass ratio. This is the first effort to train fully amortized networks for parameter estimation on real detector noise, eliminating the need for both importance sampling and likelihood assumptions. This approach allows us to realistically study the effects of real detector noise while leveraging the speed of \ac{SBI}.

This paper is organised as follows: \Cref{sec:method} describes our \ac{SBI} methods and simulations, \Cref{sec:gauss} presents results on signals generated in Gaussian noise and \Cref{sec:real} does the same on real detector noise. Finally, \Cref{sec:discuss} summarizes our findings and discusses them in the context of future developments.

\section{Methods}\label{sec:method}
The field of \ac{GW} parameter inference currently relies on Bayesian sampling methods to retrieve astrophysical information from \ac{GW} signals. These methods are all based on Bayes' theorem:
\begin{equation}
    p(\theta|x) = \frac{p(x|\theta) \cdot p(\theta)}{p(x)},
\end{equation}
where $p(\theta | x)$ is the posterior probability of the parameters $\theta$ given the observed data $x$,
$p(x | \theta)$ is the likelihood of the data $x$ given the parameters $\theta$, $p(\theta)$ is the prior of the parameters $\theta$ before observation,
and $p(x)$ is the marginal likelihood or evidence \textit{i.e.} the probability of observing the data under all considered parameter values. It is calculated as:
\begin{equation}
    p(x) = \int p(x | \theta) p(\theta) \, d\theta
    \label{eq:evidence}
\end{equation} 

\ac{SBI} is a class of Bayesian machine learning methods that utilise simulated data in order to infer probability density distributions. In this paper we use \ac{LAMPE}'s~\cite{rozet2021lampe} implementation of \ac{FMPE}~\cite{lipman2022flow,dax2023flow} as our density estimator. To train the neural network, we require only mechanistic models (in our case these are the \ac{GW} waveform models), constraints on the prior and segments of real on-site detector noise. We sample from the prior and in conjunction with our models and noise segments, we simulate synthetic data $x \sim p(x|\theta)$ to give as input to a normalising flows neural network \cite{rezende2015variational, papamakarios2021normalizing}. Normalising flows define a probability distribution $q(\theta|x)$ over $n$ number of parameters in the parameter space $\theta \in \mathbf{R}^{n}$ in terms of an invertible mapping $\psi_{x} : \mathbf{R}^{n} \rightarrow \mathbf{R}^{n}$ from a simple base distribution $q_{0}(\theta)$~\cite{rezende2015variational,papamakarios2021normalizing}:
\begin{equation}
    q(\theta|x) = (\psi_x)_* q_0(\theta) = q_0(\psi_x^{-1}(\theta)) \left| \det \frac{\partial \psi_x^{-1}(\theta)}{\partial \theta} \right|,
\end{equation} 
where $(\psi)_*$ denotes the forward flow operator, and $x$ is conditioned as $x \in \mathbf{R}^{m}$, where $m$ is the dimension of the observed data space, i.e., how many data points or features are in each observation $x$. Normalizing flows are discrete flows, such that $\psi_x$
is a collection of simpler mappings with triangular Jacobians and $\theta$ shuffling. This results in a neural density estimator, $q(\theta|x)$, that is simple to evaluate, quick to sample from and approximates the posterior $p(\theta|x)$. Flow matching is a method that uses a vector field $v_t$ to directly define the velocity of sample trajectories as they move towards the target distribution~\cite{lipman2022flow}. These trajectories are determined by solving ordinary differential equations (ODEs), which allows flow matching to achieve optimal transport without the need for discrete diffusion paths. This means that flow matching can directly reach the target distribution and compute densities more efficiently than other generative methods, such as \ac{NPSE}~\cite{sharrock2022sequential,geffner2023compositional,sohl2015deep,song2019generative,ho2020denoising}. \ac{FMPE} is a technique that applies flow matching to Bayesian inference~\cite{lipman2022flow,dax2023flow}, it works by directly aligning the estimated posterior distribution with the true posterior distribution. This alignment is achieved through a loss function that minimizes the difference between the two distributions. Due to this and the continuous nature of the flow, \ac{FMPE} can lead to a more direct and possibly more accurate estimation of the posterior as opposed to other methods such as \ac{SNPE}\cite{rezende2015variational,papamakarios2021normalizing,Bhardwaj:2023xph}. In those methods the posterior distribution is iteratively refined through sequential updates with a heavy reliance on approximations and intermediate layers, or \ac{NPSE} where they focus on estimating the score function (gradient of the log-posterior) rather than the posterior distribution itself, leading to intractable posterior densities.

In the \ac{FMPE} regime, we utilise continuous normalising flows, which are parameterised by a continuous ''time'' parameter $t \in [0,1]$, such that $q_{t=0}(\theta|x) = q_{0}(\theta)$ and $q_{t=1}(\theta|x) = q(\theta|x)$~\cite{chen2018neural}. Each $t$ defines the flow by a vector field $v_{t,x}$ on the sample space. This corresponds to the
velocity of the sample trajectories,
\begin{equation}
  \frac{d}{dt} \psi_{t,x}(\theta) = v_{t,x}(\psi_{t,x}(\theta)), \quad \psi_{0,x}(\theta) = \theta.
\end{equation}

Integrating this ODE then gives the trajectories $\theta_{t} \equiv \psi_{t,x}(\theta)$. The final density is retrieved by solving the transport equation $\partial_{t}q_{t} + div(q_{t}v_{t,x})=0$ and is: 
\begin{equation}
    q(\theta|x) = (\psi_{1,x})_* q_0(\theta) = q_0(\theta) \exp \left( -\int_0^1 \text{div} \, v_{t,x}(\theta_t) \, dt \right). 
\end{equation}

The continuous flow thus allows $v_{t,x}(\theta)$ to be specified simply by a neural network taking $\mathbf{R}^{n+m+1} \rightarrow \mathbf{R}^{n}$. The main goal of flow matching training is to make the learned vector field  $v_{t,x}$ closely follow a target vector field $u_{t,x}$. This target vector field generates a path $p_{t,x}$ that leads us towards the posterior distribution we want to estimate. By doing this, we avoid the need to solve ordinary differential equations (ODEs) during training. Although choosing the pair $(u_{t,x},p_{t,x})$ might seem complex initially,~\cite{lipman2022flow} showed that the training process becomes much simpler if we condition the path on $\theta_1$, a sample drawn from the prior distribution, instead of $x$. This is known as sample-conditional basis. For a given sample-conditional probability path $p_{t}(\theta|\theta_{1})$ with a corresponding vector field $u_{t}(\theta|\theta_{1})$, the sample-conditional flow matching loss is defined as 
\begin{align}
    L_{\text{SCFM}} =\ &\mathbb{E}_{t \sim U[0,1], x \sim p(x), \theta_1 \sim p(\theta|x), \theta_t \sim p_t(\theta_t|\theta_1)} \cdot\nonumber\\
    &\left[ \| v_{t,x}(\theta_t) - u_t(\theta_t|\theta_1) \|^2 \right].
    \label{eq:SCFM} 
\end{align}

According to~\cite{lipman2022flow}, minimising this loss is equivalent to regressing $v_{t,x}(\theta)$ on the marginal vector field $u_{t,x}$ that generates $p_{t}(\theta|x)$. Due to the sample-conditional vector field being independent of $x$ the $x-$dependence of $v_{t,x}(\theta)$ is picked up by the expectation value. Flow matching is applied to \ac{SBI} by using Bayes' theorem to make the replacement $\mathbf{E_{p(x)p(\theta|x)}} \rightarrow \mathbf{E_{p(\theta)p(x|\theta)}}$, removing the intractable expectation values, making the new \ac{FMPE} loss:
\begin{align}
    L_{\text{FMPE}} =\ &\mathbb{E}_{t \sim p(t), \theta_1 \sim p(\theta), x \sim p(x|\theta_1), \theta_t \sim p_t(\theta_t|\theta_1)} \cdot\nonumber\\
    &\left[ \| v_{t,x}(\theta_t) - u_t(\theta_t|\theta_1) \|^2 \right].
    \label{eq:FMPE}
\end{align}

We generalise the uniform distribution in \Cref{eq:SCFM} by sampling from $t \sim p(t), t \in [0, 1]$ in this expression as well. This provides more freedom to improve learning in our networks. 

The family of Gaussian sample-conditional paths are first presented in~\cite{lipman2022flow} and are given as
\begin{equation}
    p_t(\theta|\theta_1) = \mathcal{N}(\theta | \mu_t(\theta_1), \sigma_t(\theta_1)^2 \mathbf{I}_n) 
\end{equation}
where one could freely specify, contingent on boundary conditions, the time-dependant means $\mu_{t}(\theta_{1})$ and standard deviations $\sigma_{t}(\theta_{1})$. The sample-conditional probability path must be chosen to concentrate around $\theta_1$ at $t = 1$ (within a small region of size $\sigma_{min}$) in addition to being the base distribution at $t = 0$. In this work, we utilise the optimal transport paths (shown in~\cite{lipman2022flow} and used in~\cite{dax2023flow}) defined by $\mu_{t}(\theta_{1})$ and $\sigma_{t}(\theta_{1}) = 1 - (1 - \sigma_{\min}) t $ making the sample-conditional vector field have the form
\begin{equation}
    u_t(\theta|\theta_1) = \frac{\theta_1 - (1 - \sigma_{\min}) \theta}{1 - (1 - \sigma_{\min}) t}. 
\end{equation}

Training data is generated by sampling from $\theta$ from the prior and in conjunction with waveform models and detector source noise, simulating data $x$ corresponding to $\theta$. The \ac{FMPE} loss in \Cref{eq:FMPE} is minimised via empirical risk minimisation over samples $(\theta,x) \sim p(\theta)p(x|\theta)$. 

Generative diffusion or flow matching models typically handle complex, high-dimensional data (such as images) in the $\theta$ space. They often use U-Net architectures to map $\theta$ to a vector field $v(\theta)$ of the same dimension, with $t$ and an optional conditioning vector $x$ included. In the \ac{SBI} case however, and particularly in this study field, the data is often complicated whereas the parameters $\theta$ are low dimensional. This indicates that it would be more useful in our case to build the network architecture as a mapping that goes from $x$ to $v(x)$ and then apply conditioning on $\theta$ and $t$. We can therefore use any feature extraction architecture for the data and in our case we use \ac{SVD}s to extract the most informative features of the data segments. Note that \ac{SVD}s built on the \ac{GW} models may not always useful because they can remove relevant features from the noise.

\section{\label{sec:result}Results}
The results presented here were created using networks trained with simulated data from non-spinning \ac{IMBH} models using the IMRPhenomXPHM waveform approximant \cite{Pratten_2021}, generated from the 9-dimensional parameter space, injected in different kinds of noise in a single \ac{GW} detector (the LIGO Hanford detector). Furthermore, during training only the chirp-mass $\mathcal{M}$ and mass-ratio $q$ were labelled, thus limiting network's output to those two parameters. This effectively marginalises over all the other parameters (sky location, distance, phase, polarization, merger time, and inclination angle), making training easier. Time and sky-location parameters local to the (single) detector were used instead of the standard geocenter time, right-ascension and declination parameters. This enables faster training while still allowing us to investigate the effect of real noise on the inference.

For this exploratory investigation the priors used are listed in \Cref{tab:parameters}. The high masses allow for a segment length of \SI{0.5}{\second} and a high frequency cutoff of \SI{256}{\hertz}, while the low frequency cutoff is set to the usual \SI{20}{\hertz}, relevant for the Advanced LIGO zero-detuned high power~\cite{PSD} noise curve used to generate Gaussian noise. We do highlight that for GW190521 which has similar mass parameters as our injection for this exploratory investigation, previous Bayesian analyses have gone below this threshold \cite{LIGOScientific:2020ufj}, while others have also used \SI{20}{\hertz}, e.g. \cite{LIGOScientific:2021usb}.

\begin{table}[ht]
    \centering
    \begin{tabular}{l l l}
        \text{Parameter} & \textbf{Range} & \textbf{Prior} \\
        \hline
        Chirp mass & 80-120 $M_\odot$ & Uniform \\
        Mass ratio & 0.3-1.0 & Uniform \\
        Luminosity distance & 1-4 Gpc & Uniform \\
        Time & 20 ms wide & Uniform \\
    \end{tabular}
    \caption{Prior parameters. All other parameters are using the standard uniform priors~\cite{gwtc3}.}
    \label{tab:parameters}
\end{table}

This work uses \ac{LAMPE}'s implementation of \ac{FMPE} which we combine with GWpy~\cite{macleod2021gwpy} and Bilby library~\cite{ashton2019bilby,bilby_pipe_paper}. In particular, we are using a Multi-Layer Perceptron with 32 layers of 256 features and Exponential Linear Unit activation function. Despite the risks of vanishing gradients, those hyper parameters did provide the best results over a sweep from $2^{3-6}$ layers and $2^{6-10}$ features. Training was done using about 10 million segments in the frequency domain, each including a simulated waveform randomly drawn from the prior, taking about 1 day on a A100 GPU.

\subsection{\label{sec:gauss}Gaussian noise}

\begin{figure*}
    \centering
    \includegraphics[width=\textwidth]{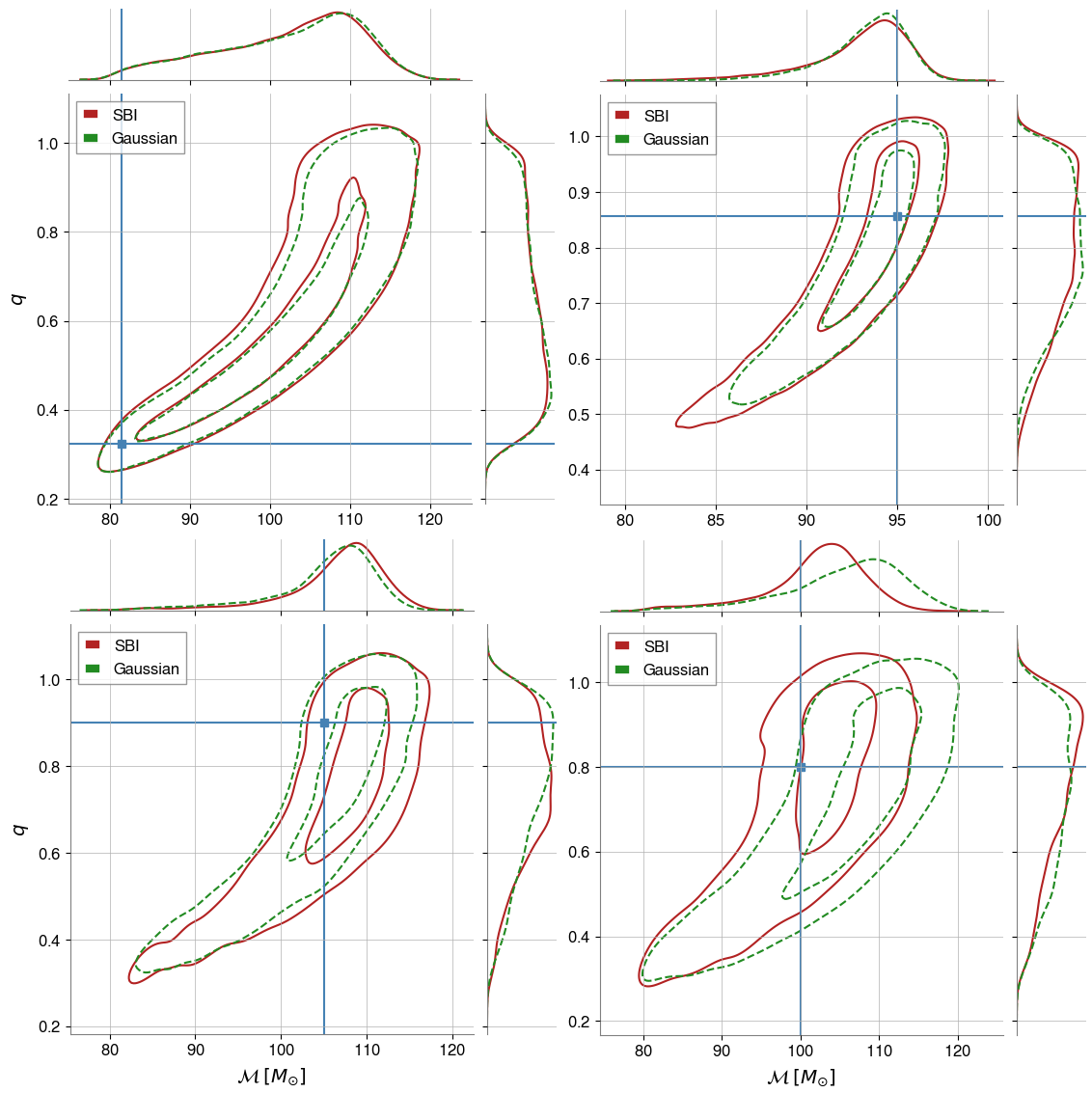}
    \caption{\ac{PDF}s for the chirp-mass $\mathcal{M}$ and the mass-ratio $q$ from sampling a Gaussian Likelihood and a \ac{SBI} network trained with (top panels) Gaussian noise on an injection into Gaussian noise, and trained with (bottom panels) real detector noise on an injection into real detector noise. The blue lines show the injected values.}
    \label{fig:combined_corner}
\end{figure*}

When trained on injections with Gaussian noise, we expect for the \ac{SBI} network to converge towards a representation of the conditional posterior mapping able to directly sample from the posterior with the correspondingly estimated Gaussian likelihood. To check the convergence of the training we perform a set of simulated injections in Gaussian noise in the LIGO Hanford detector using IMRPhenomXPHM and recover them with both the trained \ac{SBI} network, as well as the Dynesty~\cite{speagle2020dynesty} sampler implemented in the bilby software library. We used the same Advanced LIGO zero-detuned high power~\cite{PSD} noise curve and the IMRPhenomXPHM waveform model. We ran 4 parallel analyses using 4000 live points each. Two examples are shown in the top panels of \Cref{fig:combined_corner}, where the sampled distributions are near-identical with \ac{JSD} values of 0.001 nat and 0.006 nat, respectively, thus confirming that our training samples accurately represented the 9-dimensional parameter space. We provide the \ac{JSD} values as they are commonly used in gravitational-wave astronomy~\cite{PhysRevX.9.031040}. A typical threshold to indicate statistical agreement is 0.035 nat.

\begin{figure}
    \centering
    \includegraphics[width=\linewidth]{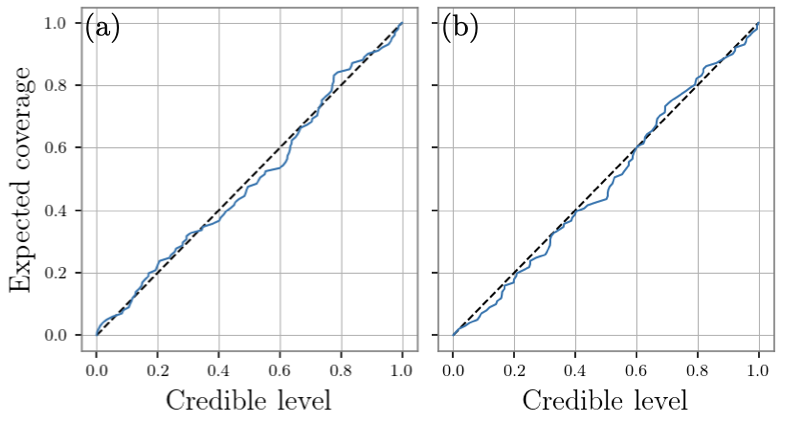}
    \caption{Percentile-percentile plots on the multi-dimensional PDF sampled from the converged neural network trained on a) Gaussian noise (p-value 0.64) and b) real detector noise (p-value 0.66), using 100 injections.}
    \label{fig:pp_plots}
\end{figure}

Furthermore, a 2-dimensional percentile-percentile test (\Cref{fig:pp_plots}, panel (a)) on 100 injections recovered with the trained network shows that our credible intervals behave as expected, which matches the performance from stochastic sampling, see for instance \cite{Romero-Shaw:2020owr}. We calculated the two-dimensional mass coverage with a Monte Carlo integration \cite{hermans2022trustcrisissimulationbasedinference} for this plot.

\subsection{\label{sec:real}LIGO detector noise}
The training data was generated by injecting simulated signals at randomly selected starting points within a roughly \SI{14}{\hour} stretch of LIGO Hanford data around February 20$^{th}$ 2020 (GPS time 1266213786.7). For training, we used the same network architecture from~\Cref{sec:gauss}. In most cases this reproduced the results from the Gaussian likelihood as sampled with Dynesty using bilby which estimated the \ac{PSD} using the default median average method settings~\cite{bilby_pipe_paper}, other settings being identical to those in the previous section. For example, the results in the bottom right panel of Figure 1 reached a JSD of 0.004 nat and 0.013 nat for the mass ratio $q$ and chirp-mass $\mathcal{M}$, respectively.

However, for a fraction of injections, by not assuming Gaussianity nor stationarity, the network outperformed the Gaussian likelihood, as shown in~\Cref{fig:combined_corner}, bottom left panel, with a \ac{JSD} of 0.008 nat for the mass ratio $q$, and 0.07 nat for the chirp-mass $\mathcal{M}$.

The statistical reliability of the credible interval of this network was assessed with a percentile-percentile plot (see~\Cref{fig:pp_plots} panel (b)), which gave the expected diagonal behaviour. Note that the network is by construction specific to the data used in training, and as such is only amortized over observations in noise with similar properties. In practice it may be necessary to retrain the network if the detector characterizations analyses find differences.

\section{Discussion}
\label{sec:discuss}

The improvement achieved by modelling the real noise distribution in our results is comparable to the Score-Based Likelihood Characterization findings of \cite{2023mla..confE..17L}. However, unlike their approach, our trained \ac{SBI} network directly generates samples from the target conditional posterior distribution and does not require an extra sampling step. On this limited example, after a $\approx \SI{24}{\hour}$ training on a A100 GPU, the bottleneck for inference time shifts to our GPU's I/O capacities, reducing it to a few milliseconds, whereas standard sampling requires several hours.

Our results thus represent a first look at how simulation-based inference may be able to perform optimal parameter estimation using real detector noise.
Current limitations include the need to characterize and quantify noise features beyond Gaussianity and stationarity. We emphasise however that this works lays a strong foundation and can in future investigations be used to focus on precessing-spin analyses that also include calibration error modeling. Testing on simulated analytical non-Gaussian noise distributions, and testing with different \ac{PSD} estimation methods such as BayesWave~\cite{PhysRevD.103.044006}, will be able to lead us to robust \ac{SBI} architectures for \ac{GW} inference.

Beyond the extension to precessing systems, future studies will also involve increasing the domain of applicability, specifically lowering the mass range on which the network is trained, as currently the analysis would not be reliable on lower-mass signals, and including multiple detectors and eccentric signals. However, given our \ac{SVD} compression pre-processing step, and the performance of other compression techniques on GW signals, for instance Reduced-Order-Modelling methods (see~\cite{ROM_review} for a review), the scaling is expected to be manageable. Additionally, work on marginalising over multiple waveform approximants such that the resulting 
marginalised posterior distribution encompasses errors from waveform systematics and will help towards enlarging the domain of applicability of the networks. 

\section*{Acknowledgements}

The authors would like to thank Stephen Green, Maximilian Dax, Virginia d'Emilio and Alex Nitz for helpful discussions. This work was supported by the UK's Science and Technology Facilities Council grant ST/V005618/1, the Royal Society Award ICA\textbackslash R1\textbackslash 231114 and the Leverhulme Trust Fellowship IF-2024-038.
This research has made use of data or software obtained from the Gravitational Wave Open Science Center (gwosc.org), a service of the LIGO Scientific Collaboration, the Virgo Collaboration, and KAGRA. This material is based upon work supported by NSF's LIGO Laboratory which is a major facility fully funded by the National Science Foundation, as well as the Science and Technology Facilities Council (STFC) of the United Kingdom, the Max-Planck-Society (MPS), and the State of Niedersachsen/Germany for support of the construction of Advanced LIGO and construction and operation of the GEO600 detector. Additional support for Advanced LIGO was provided by the Australian Research Council. Virgo is funded, through the European Gravitational Observatory (EGO), by the French Centre National de Recherche Scientifique (CNRS), the Italian Istituto Nazionale di Fisica Nucleare (INFN) and the Dutch Nikhef, with contributions by institutions from Belgium, Germany, Greece, Hungary, Ireland, Japan, Monaco, Poland, Portugal, Spain. KAGRA is supported by Ministry of Education, Culture, Sports, Science and Technology (MEXT), Japan Society for the Promotion of Science (JSPS) in Japan; National Research Foundation (NRF) and Ministry of Science and ICT (MSIT) in Korea; Academia Sinica (AS) and National Science and Technology Council (NSTC) in Taiwan.
The authors are grateful for computational resources provided by the LIGO Laboratory and Cardiff University and supported by National Science Foundation Grants PHY-0757058 and PHY-0823459, and STFC grants ST/I006285/1 and ST/V005618/1.

\section*{Data Availability}

The data utilized for this work will be freely available upon reasonable request to the corresponding author.



\bibliographystyle{custom_ieetr}
\bibliography{apssamp} 

\providecommand{\noopsort}[1]{}\providecommand{\singleletter}[1]{#1}%
\begin{thebibliography}{10}

\bibitem{PhysRevD.102.062003}
A.~Buikema, C.~Cahillane, G.~L. Mansell, {\em et~al.} {\em Phys. Rev. D},
  vol.~102, p.~062003, Sep 2020.

\bibitem{PhysRevLett.123.231108}
F.~Acernese, M.~Agathos, L.~Aiello, {\em et~al.} {\em Phys. Rev. Lett.},
  vol.~123, p.~231108, Dec 2019.

\bibitem{10.1093/ptep/ptaa125}
T.~Akutsu, M.~Ando, K.~Arai, {\em et~al.} {\em Progress of Theoretical and
  Experimental Physics}, vol.~2021, p.~05A101, 08 2020.

\bibitem{gwtc3}
R.~Abbott, T.~D. Abbott, F.~Acernese, {\em et~al.} {\em Phys. Rev. X}, vol.~13,
  p.~041039, Dec 2023.

\bibitem{annurev:/content/journals/10.1146/annurev-astro-032620-021835}
J.~E. Greene, J.~Strader, and L.~C. Ho {\em Annual Review of Astronomy and
  Astrophysics}, vol.~58, no.~Volume 58, 2020, pp.~257--312, 2020.

\bibitem{ashton2019bilby}
G.~Ashton, M.~H{\"u}bner, P.~D. Lasky, {\em et~al.} {\em The Astrophysical
  Journal Supplement Series}, vol.~241, no.~2, p.~27, 2019.

\bibitem{powell2018}
J.~Powell {\em Class. Quantum Grav.}, vol.~35, p.~155017, Aug. 2018.

\bibitem{davis2022}
D.~Davis, T.~B. Littenberg, I.~M. {Romero-Shaw}, {\em et~al.} {\em Class.
  Quantum Grav.}, vol.~39, p.~245013, Dec. 2022.

\bibitem{PhysRevD.106.042006}
S.~Hourihane, K.~Chatziioannou, M.~Wijngaarden, {\em et~al.} {\em Phys. Rev.
  D}, vol.~106, p.~042006, Aug 2022.

\bibitem{PhysRevD.105.103021}
R.~Macas, J.~Pooley, L.~K. Nuttall, {\em et~al.} {\em Phys. Rev. D}, vol.~105,
  p.~103021, May 2022.

\bibitem{10.1093/mnras/stad341}
G.~Ashton {\em Monthly Notices of the Royal Astronomical Society}, vol.~520,
  pp.~2983--2994, 02 2023.

\bibitem{cranmer2020frontier}
K.~Cranmer, J.~Brehmer, and G.~Louppe {\em Proceedings of the National Academy
  of Sciences}, vol.~117, no.~48, pp.~30055--30062, 2020.

\bibitem{green2020gravitational}
S.~R. Green, C.~Simpson, and J.~Gair {\em Physical Review D}, vol.~102, no.~10,
  p.~104057, 2020.

\bibitem{delaunoy2020lightning}
A.~Delaunoy, A.~Wehenkel, T.~Hinderer, {\em et~al.} {\em arXiv preprint
  arXiv:2010.12931}, 2020.

\bibitem{dax2021real}
M.~Dax, S.~R. Green, J.~Gair, {\em et~al.} {\em Physical review letters},
  vol.~127, no.~24, p.~241103, 2021.

\bibitem{dax2022group}
M.~Dax, S.~Green, J.~Gair, {\em et~al.}, ``Group equivariant neural posterior
  estimation,'' in {\em ICLR 2022}, 2022.

\bibitem{dax2023flow}
M.~Dax, J.~Wildberger, S.~Buchholz, {\em et~al.} {\em arXiv preprint
  arXiv:2305.17161}, 2023.

\bibitem{Bhardwaj:2023xph}
U.~Bhardwaj, J.~Alvey, B.~K. Miller, {\em et~al.} {\em ArXiV}, 4 2023.

\bibitem{alvey2024simulation}
J.~Alvey, U.~Bhardwaj, V.~Domcke, {\em et~al.} {\em Physical Review D},
  vol.~109, no.~8, p.~083008, 2024.

\bibitem{wildberger2023adapting}
J.~Wildberger, M.~Dax, S.~R. Green, {\em et~al.} {\em Physical Review D},
  vol.~107, no.~8, p.~084046, 2023.

\bibitem{legin2023framework}
R.~Legin, Y.~Hezaveh, L.~Perreault-Levasseur, and B.~Wandelt {\em The
  Astrophysical Journal}, vol.~943, no.~1, p.~4, 2023.

\bibitem{2023mla..confE..17L}
R.~{Legin}, K.~{Wong}, M.~{Isi}, {\em et~al.}, ``{Towards Unbiased
  Gravitational-Wave Parameter Estimation using Score-Based Likelihood
  Characterization},'' in {\em Machine Learning for Astrophysics}, p.~17, July
  2023.

\bibitem{rozet2021lampe}
F.~Rozet, A.~Delaunoy, B.~Miller, {\em et~al.}, ``{LAMPE}: Likelihood-free
  amortized posterior estimation,'' 2021.

\bibitem{lipman2022flow}
Y.~Lipman, R.~T. Chen, H.~Ben-Hamu, {\em et~al.} {\em arXiv preprint
  arXiv:2210.02747}, 2022.

\bibitem{rezende2015variational}
D.~Rezende and S.~Mohamed, ``Variational inference with normalizing flows,'' in
  {\em International conference on machine learning}, pp.~1530--1538, PMLR,
  2015.

\bibitem{papamakarios2021normalizing}
G.~Papamakarios, E.~Nalisnick, D.~J. Rezende, {\em et~al.} {\em Journal of
  Machine Learning Research}, vol.~22, no.~57, pp.~1--64, 2021.

\bibitem{sharrock2022sequential}
L.~Sharrock, J.~Simons, S.~Liu, and M.~Beaumont {\em arXiv preprint
  arXiv:2210.04872}, 2022.

\bibitem{geffner2023compositional}
T.~Geffner, G.~Papamakarios, and A.~Mnih, ``Compositional score modeling for
  simulation-based inference,'' in {\em International Conference on Machine
  Learning}, pp.~11098--11116, PMLR, 2023.

\bibitem{sohl2015deep}
J.~Sohl-Dickstein, E.~Weiss, N.~Maheswaranathan, and S.~Ganguli, ``Deep
  unsupervised learning using nonequilibrium thermodynamics,'' in {\em
  International conference on machine learning}, pp.~2256--2265, PMLR, 2015.

\bibitem{song2019generative}
Y.~Song and S.~Ermon {\em Advances in neural information processing systems},
  vol.~32, 2019.

\bibitem{ho2020denoising}
J.~Ho, A.~Jain, and P.~Abbeel {\em Advances in neural information processing
  systems}, vol.~33, pp.~6840--6851, 2020.

\bibitem{chen2018neural}
R.~T. Chen, Y.~Rubanova, J.~Bettencourt, and D.~K. Duvenaud {\em Advances in
  neural information processing systems}, vol.~31, 2018.

\bibitem{Pratten_2021}
G.~Pratten, C.~García-Quirós, M.~Colleoni, {\em et~al.} {\em Physical Review
  D}, vol.~103, May 2021.

\bibitem{PSD}
B.~P. Abbott, R.~Abbott, T.~D. Abbott, {\em et~al.} {\em Living Reviews in
  Relativity}, vol.~23, no.~1, p.~3, 2020.

\bibitem{LIGOScientific:2020ufj}
R.~Abbott {\em et~al.} {\em Astrophys. J. Lett.}, vol.~900, no.~1, p.~L13,
  2020.

\bibitem{LIGOScientific:2021usb}
R.~Abbott {\em et~al.} {\em Phys. Rev. D}, vol.~109, no.~2, p.~022001, 2024.

\bibitem{macleod2021gwpy}
D.~M. Macleod, J.~S. Areeda, S.~B. Coughlin, {\em et~al.} {\em SoftwareX},
  vol.~13, p.~100657, 2021.

\bibitem{bilby_pipe_paper}
I.~M. Romero-Shaw {\em et~al.} {\em Mon. Not. Roy. Astron. Soc.}, vol.~499,
  no.~3, pp.~3295--3319, 2020.

\bibitem{speagle2020dynesty}
J.~S. Speagle {\em Monthly Notices of the Royal Astronomical Society},
  vol.~493, no.~3, pp.~3132--3158, 2020.

\bibitem{PhysRevX.9.031040}
B.~P. Abbott, R.~Abbott, T.~D. Abbott, {\em et~al.} {\em Phys. Rev. X}, vol.~9,
  p.~031040, Sep 2019.

\bibitem{Romero-Shaw:2020owr}
I.~M. Romero-Shaw {\em et~al.} {\em Mon. Not. Roy. Astron. Soc.}, vol.~499,
  no.~3, pp.~3295--3319, 2020.

\bibitem{hermans2022trustcrisissimulationbasedinference}
J.~Hermans, A.~Delaunoy, F.~Rozet, {\em et~al.}, ``A trust crisis in
  simulation-based inference? your posterior approximations can be
  unfaithful,'' 2022.

\bibitem{PhysRevD.103.044006}
N.~J. Cornish, T.~B. Littenberg, B.~B\'ecsy, {\em et~al.} {\em Phys. Rev. D},
  vol.~103, p.~044006, Feb 2021.

\bibitem{ROM_review}
M.~Tiglio and A.~Villanueva {\em Living Reviews in Relativity}, vol.~25, no.~1,
  p.~2, 2022.

\end{thebibliography}





\end{document}